\documentclass[letterpaper, 10 pt, conference]{ieeeconf}  

\IEEEoverridecommandlockouts                              

\usepackage{kpfonts}
\usepackage{times}
\usepackage{amsmath} 
\usepackage{amssymb}  
\usepackage{mathrsfs} 
\usepackage{graphicx,cite}
\usepackage{enumerate}
\usepackage{comment}
\usepackage{algorithm}
\usepackage{algpseudocode}
\usepackage{tcolorbox}
\usepackage{tabularx}
\usepackage{array}
\usepackage{dsfont}
\usepackage{setspace}
\usepackage{caption} 
\usepackage{tikz}
\usetikzlibrary{automata, positioning}

\newtheorem{rem}{Remark}

\newtheorem{lem}{Lemma}
\newtheorem{cor}{Corollary}

\newtheorem{defn}{Definition}

\newcommand{\remend}{\relax\ifmmode\else\unskip\hfill\fi\hbox{$\bullet$}}

\providecommand{\keywords}[1]{\textbf{\textit{Index terms---}} #1}

\title{\LARGE \bf
Distributionally Robust Safety Verification for Markov Decision Processes}

\author{Abhijit Mazumdar, Yuting Hou and Rafal Wisniewski, \IEEEmembership{Senior Member, IEEE}
\thanks{This work has been supported by the Independent Research Fund Denmark (DFF) under Project SAFEmig (Project Number 3105-00173B).}
\thanks{All authors are with the Section of Automation $\&$ Control, Aalborg University, 9220 Aalborg East, Denmark (e-mail: \{abma, yutingh, raf\}@es.aau.dk). Corresponding author: Abhijit Mazumdar.}
}

\graphicspath{{figs/}}

\begin{document}

\maketitle
\thispagestyle{empty}
\pagestyle{empty}

\begin{abstract}

   In this paper, we propose a distributionally robust safety verification method for Markov decision processes where only an ambiguous transition kernel is available instead of the precise transition kernel. We define the ambiguity set around the nominal distribution by considering a Wasserstein distance. To this end, we introduce a \textit{robust safety function} to characterize probabilistic safety in the face of uncertain transition probability. First, we obtain an upper bound on the robust safety function in terms of a distributionally robust Q-function. Then, we present a convex program-based distributionally robust Q-iteration algorithm to compute the robust Q-function. By considering a numerical example, we demonstrate our theoretical results.   
  \end{abstract}
  \vspace{0cm}
\section{Introduction}
Safety verification is vital to safety-critical systems, where unmodeled uncertainties and unpredictable variations can pose serious risks. Traditional approaches to probabilistic safety verification often assume that the system’s behavior can be precisely characterized by known probability distributions. However, accurately identifying these probability distributions in practical scenarios is often challenging due to limited data, environmental variability, or complex dynamics. Consequently, any safety guarantees based on approximate or estimated probability models may become invalid if there is a mismatch between the assumed and actual distributions governing the system's transitions.
\par In stochastic optimization, distributionally robust optimization (DRO) has emerged as a powerful tool for providing reliable guarantees under model uncertainty. DRO frameworks account for this uncertainty by optimizing for the worst-case distribution within a specified ambiguity set—a collection of probability distributions that captures possible variations around the nominal model. By considering the worst-case scenario within this set, DRO ensures that the safety guarantees remain robust, even when the true distribution deviates from the computed or assumed one. This approach allows safety-critical systems to operate with greater resilience to uncertainties in their underlying probabilistic models. There are many approaches used in DRO such as Wasserstein distance-based \cite{mohajerin2018data,gao2023distributionally,yu2024fast}, moment-based \cite{hao2021moment}, $\Phi$-divergence-based \cite{wu2024understanding,love2014classification}, etc. 
\\

\textit{Related Literature:}
The concept of safety that we consider in this work is the so-called $p$-safe. It has been developed in \cite{wisniewski2020p} and extensively studied in \cite{wisniewski2021safety, bujorianu2021stochastic, bujorianu2021p, bujorianu2020stochastic}. A system is called to be $p$-safe if the system states do not visit the dangerous states before reaching the control goal with a probability more than a given threshold $p$. In \cite{wisniewski2023probabilistic}, considering a Markov decision process (MDP), the safety function is reformulated by expected cost and a recursive expression is presented using dynamic programming problem.
\par In solving DRO optimization, the ambiguity set plays an important role. Wasserstein metric, which is a distance on the space of probability distributions, has been widely used to construct ambiguity set. However, use of Wasserstein distance makes the original problem an infinite-dimensional optimization over probability measures. By using Kantorovich duality \cite{mohajerin2018data,kantorovich1958space} resolves the infinite-dimension issue in the worst-case problem and represents the worse-case problem by a finite-dimensional convex optimization. In the context of an MDP, \cite{iyengar2005robust} introduces a distributionally robust Bellman equation based on which, a robust value iteration algorithm and a robust policy iteration algorithm are developed. Considering Wasserstein ambiguity set, \cite{yang2017convex} presents a method for finding the robust optimal policy. To carry out safety assessment, \cite{yang2018dynamic} uses moment-based ambiguity to define the uncertain distribution, enabling safety verification under partially known disturbance distributions. 
\\
\\
\textit{Our Contributions:} 
In this paper, we consider an MDP with a known nominal transition probability and an ambiguity set. We aim to investigate probabilistic safety subject to imperfect knowledge about the true transition probability. Specifically, our goal is to compute the worst case probability of reaching an unsafe state of the MDP before visiting a goal state. To do so, we formulate the problem as a distributionally robust sequential optimization. We use Wasserstein distance to construct the ambiguity set \cite{mohajerin2018data}. To characterize distributionally robust safety, we  introduce the robust safety function which is the worst case probability of reaching an unsafe state before visiting a goal state. We observe that the robust safety function is the solution of an infinite-dimensional DRO with Wasserstein ambiguity set. To convert the problem into a finite-dimensional convex program, we first observe that the robust safety function has the form of a value function. This allows us to find an upper bound of the robust safety function in terms of the robust Q-function. Then we derive a convex program for distributionally robust Q-function. Based on this convex program, we present a model-based distributionally robust Q-iteration algorithm to estimate an upper bound of the robust safety function.
\par The following are our main contributions:
\begin{enumerate}[i)]
    \item We extend the notion of probabilistic safety namely $p$-safety to its distributionally robust counterpart in the context of MDPs.
    \item We obtain an upper bound for robust safety function in terms of a distributionally robust Q-function for MDPs.
    \item We develop a finite convex program formulation of the distributionally Q-function. Subsequently, we present a model-based robust Q-iteration algorithm for estimating an upper bound for the robust safety function.
\end{enumerate}

The organization of the rest of the paper is as follows. In section \ref{problem_formulation}, we introduce the MDP and various definitions. We introduce Wasserstein metric to describe the distance of two probability distributions and use Katorovich’s duality to reformulate it. Then, we formally define robust $p$-safety. We present the main results of the paper in Section \ref{DR safety verification}.
We provide an upper bound for the robust safety function in terms of the distributionally robust Q-function. The convex formulation of the robust Q-function is derived. Based on this result, we develop the distributionally robust Q-iteration, given the nominal distribution. In section \ref{numerical results}, we demonstrate with an example how the value of the robust safety function changes as the radius of the Wasserstein ball increases. Finally, we present the concluding remark and our future work plan in section \ref{sec_conclusion}.
\\
\\
\textit{Notations:}
For any set $S$, we use $\mathscr{M}(S)$ to denote the set of all Borel probability measure over $S$. For any real number $x$, by $|x|$ we denote its absolute value, while for a set $S$, $|S|$ denotes its cardinality. For any probability measure $\mu$ over a discrete set $X$, we use $L^1(\mu)$ to denote all functions that satisfies $\sum_{x\in X} |f(x)|\mu(x)<\infty$. 
\\
\\
\section{Background and Problem Formulation} \label{problem_formulation}
We consider a Markov decision process (MDP) with a set of finitely many states $\mathcal{X}$, a set of finitely many actions $\mathcal{A}$. We partition the set of state $\mathcal{X}$ into three subsets: a set of all \textit{goal or target} states $E$, a set of \textit{forbidden or unsafe} states $U$, and a \textit{taboo} set $H:=\mathcal{X}\setminus (E\cup U)$. In our setup, the goal set $E$ and the forbidden set $U$ are terminal, meaning the process gets terminated at those states. We consider a sample space $\Omega$ of the form $\omega=(x_0,a_0,x_1,a_1,\ldots)\in(\mathcal{X}\times\mathcal{A})^{\infty}$ with $x_i\in\mathcal{X}$ and $a_i\in\mathcal{A}$. The sample space $\Omega$ is equipped with the $\sigma$-algebra $\mathcal{F}$ generated by coordinate mappings: $X_t(w)=x_t$ and $A_t(w)=a_t$. With a slight abuse of notation, we use upper case $X_t$ and $A_t$ to denote random variables, while $x_t$ and $a_t$ represent their deterministic values, or realizations, at time step $t$. We use $\pi(a|x)$ to denote a stationary policy where $\pi: \mathcal{X}\rightarrow \mathscr{M}(\mathcal{A})$.
From any state $x\in H$, the MDP makes a transition to the next state $y\in \mathcal{X}$ with a probability $P_{x,a}(y)$. We define the following set $\mathcal{P}_{x,a}:= \{ P_{x,a}(y) \}_{y\in \mathcal{X}}$ and $\mathscr{P}:= \{ \mathcal{P}_{x,a} \}_{(x,a)\in H\times \mathcal{A}}$. Throughout the paper, we assume that the transition probabilities are time-invariant. We use $\mathds{P}_{\pi}^{\mathscr{P}}[.]$ and $\mathds{E}_{\pi}^{\mathscr{P}}[.]$ to denote the probability measure and the expectation with respect to a fixed policy $\pi$ and transition probability $\mathscr{P}$, respectively. 
\par If the transition probability $\mathscr{P}$ is precisely known, then probabilistic safety can be characterized by defining the \textit{safety function} \cite{wisniewski2023probabilistic}.
\begin{defn}[Safety function] \cite{wisniewski2023probabilistic}
    Consider a fixed policy $\pi$ and a fixed initial state $x\in H$. The safety function $S^{{\mathscr{P}}}_{\pi}(x)$ is defined as the probability that any realization starting from the initial state $x$ hits the unsafe set $U$ before hitting the goal set $E$ following policy $\pi$, i.e.,
    \begin{equation*}
        S^{{\mathscr{P}}}_{\pi}(x) := \mathds{P}_{\pi}^{\mathscr{P}}[\tau_U < \tau_E\big| X_0=x],
    \end{equation*}
    where $\tau_S$ is the first hitting time of a set $S$. \remend
\end{defn} 
 \par In reality, we may not have access to the precise transition probabilities of the MDP. Instead, we only have a nominal transition probability $\mathscr{P}$ with a Wasserstein-type ambiguity set $\mathscr{D}^{\delta}$ as given below \cite{mohajerin2018data}:
 \begin{equation*}
\begin{split}
    & \mathscr{D}^{\delta} := \varprod_{(x,a)\in H\times \mathcal{A}} \mathcal{D}^{\delta}_{x,a} \ , \\
    & \mathcal{D}^{\delta}_{x,a} := \Big{\{}\tilde{\mathcal{P}}_{x,a}\in \mathscr{M}(\mathcal{X}) \ \big| \ W(\tilde{\mathcal{P}}_{x,a},\mathcal{P}_{x,a}) \leq \delta  \Big{\}},
    \end{split}
\end{equation*}
 where  $W(\tilde{\mathcal{P}}_{x,a},\mathcal{P}_{x,a})$ is the $1$-Wasserstein distance and is of the following form.
     \begin{equation*}
    \begin{split}
    & W(\tilde{\mathcal{P}}_{x,a},\mathcal{P}_{x,a}) \\
    & :=  \Big{\{} \underset{\Gamma\in \mathscr{M}(\mathcal{X} \times \mathcal{X})}{\text{min }} \ \sum_{(y,z)\in \mathcal{X} \times \mathcal{X}} \Gamma(y,z) |y-z| \\
    &  \text{s.t.} \ \sum_{z\in \mathcal{X}} \Gamma(y,z)= \tilde{{P}}_{x,a}(y), \ \sum_{y\in \mathcal{X}} \Gamma(y,z)= {P}_{x,a}(z) \Big{\}}
\end{split}
\end{equation*}
The Wasserstein distance can be represented in a dual form using Kantorovich's duality as follows \cite{gao2023distributionally}: 
\small
\begin{equation*}
\begin{split}
&W(\tilde{\mathcal{P}}_{x,a},\mathcal{P}_{x,a})\\
& = \underset{f\in {L}^1(\tilde{P}_{x,a}), g\in {L}^1({P}_{x,a})}{\text{sup}}
   \begin{cases}
    \sum_{z_1\in \mathcal{X}} f(z_1) \tilde{{P}}_{x,a}(z_1) + \sum_{z_2\in \mathcal{X}} g(z_2) {P}_{x,a}(z_2) \\
    s.t. \ f(z_1) + g(z_2) \leq |z_1-z_2|, \ \forall z_1,z_2 \in \mathcal{X}
    \end{cases} 
  \end{split}
\end{equation*}
\normalsize
Then, we get the following expression for the Wasserstein distance $W(\tilde{\mathcal{P}}_{x,a},\mathcal{P}_{x,a})$:
\small
\begin{equation}
\begin{split}
& W(\tilde{\mathcal{P}}_{x,a},\mathcal{P}_{x,a})\\
    & = \underset{f\in {L}^1(\tilde{P}_{x,a}), g\in {L}^1({P}_{x,a})}{\text{sup}}
   \begin{cases}
    \sum_{z_1\in \mathcal{X}} f(z_1) \tilde{{P}}_{x,a}(z_1) + \sum_{z_2\in \mathcal{X}} g(z_2) {P}_{x,a}(z_2) \\
    s.t. \  g(z_2) \leq \underset{z_1}{\text{inf}} \ \big( |z_1-z_2| - f(z_1) \big) , \ \forall z_2 \in \mathcal{X} 
    \end{cases} \\
    & = \underset{f\in {L}^1(\tilde{P}_{x,a})}{\text{sup}} \ 
    \sum_{z_1\in \mathcal{X}} f(z_1) \tilde{{P}}_{x,a}(z_1) + \sum_{z_2\in \mathcal{X}} \underset{z_1}{\text{inf}} \ \big( |z_1-z_2| - f(z_1) \big) {P}_{x,a}(z_2) 
  \end{split}
  \label{Wasserstein_dual2}
\end{equation}
\normalsize
\par To characterize safety with ambiguous transition probability, we now introduce the distributionally robust safety function. 
\begin{defn}[Robust safety function]
    Consider a fixed policy $\pi$ and a given nominal transition probability $\mathscr{P}$ with an ambiguity set $\mathscr{D}^{\delta}$. We define the \text{distributionally robust safety function} or simply \text{robust safety function} $S^{\delta,\mathscr{P}}_{\pi}(x)$ as the worst-case probability that any realization starting from $X_0=x\in H$ visits the forbidden set $U$ before visiting the goal set $E$, i.e.,
    \begin{equation*}
        S^{\delta,\mathscr{P}}_{\pi}(x):= \underset{\tilde{\mathscr{P}}\in \mathscr{D}^{\delta}}{\text{sup}} \ \mathds{P}_{\pi}^{\tilde{\mathscr{P}}}[\tau_U < \tau_E \big| X_{0}=x] 
        \label{dr_safety_fn}
    \end{equation*}
    \remend
\end{defn}
\par We consider the following notion of distributionally robust probabilistic safety, namely robust $p$-safety. This definition is inspired by the $p$-safety notion for standard MDPs introduced in \cite{wisniewski2023probabilistic}.  
\begin{defn}[Robust $p$-safety] 
    Suppose policy $\pi$, robustness parameter $\delta$ and the safety parameter $p\in (0,1)$ are fixed. A state $x\in H$ is called robust $p$-safe if $S^{\delta,\mathscr{P}}_{\pi}(x)\leq p$. We call an MDP to be robust $p$-safe if $\underset{x\in H}{\text{max}}\ S^{\delta,\mathscr{P}}_{\pi}(x)\leq p$. \remend
\end{defn}
\par Our goal is to assess the robust $p$-safety of an MDP for a \textit{evaluation policy} $\pi$. The nominal transition probability $\mathscr{P}$, robustness parameter $\delta$, and a safety parameter $p$ are given. The robust safety function, as presented in \eqref{dr_safety_fn}, is, in general, intractable to compute as it turns out to be an infinite-dimensional optimization over probability measure \cite{mohajerin2018data}. However, we aim to calculate the safety function in a tractable manner by converting the problem into a finite-dimensional dual problem. To this end, we use Kantorovich duality for Wasserstein distance.
\\
\\
\\
\section{Distributionally robust safety verification}\label{DR safety verification}
In this section, we present all the main results to evaluate robust $p$-safety.
 \par We express the robust safety function as a worst-case finite horizon cost function as shown below.
  \begin{lem} \label{dr_safe_kappa}
      The robust safety function $S^{\delta,\mathscr{P}}_{\pi}(x)$ can be expressed as follows:
      \begin{equation*}
         S^{\delta,\mathscr{P}}_{\pi}(x) = \underset{\tilde{\mathscr{P}}\in \mathscr{D}^{\delta}}{\text{sup}} \ \mathds{E}_{\pi}^{\tilde{\mathscr{P}}} \sum_{t=0}^{\tau-1} \Big[ \kappa^{\tilde{{\mathscr{P}}}}(X_t, A_t) \big| X_0=x \Big],
      \end{equation*}
      where, $\tau = \tau_{E\cup U}$ and $\kappa^{\tilde{{\mathscr{P}}}}(x,a)= \sum_{y \in U} \tilde{{P}}_{x,a}(y)$.
  \end{lem}
  \begin{proof}
The safety function $S^{\mathscr{P}}_{\pi}(x)$ can be expressed as follows \cite{wisniewski2023probabilistic},
\begin{equation*}
    S^{\mathscr{P}}_{\pi}(x)= \mathds{E}_{\pi}^{{\mathscr{P}}} \Big[ \sum_{t=0}^{\tau-1} \kappa^{{\mathscr{P}}}(X_t, A_t) \big| X_0=x  \Big],
\end{equation*}
 where, $\kappa^{{\mathscr{P}}}(x,a)= \sum_{y \in U} {{P}}_{x,a}(y)$.
  \par Now, from the definition of the robust safety function, 
  \small
  \begin{equation*}
    \begin{split}
        & S^{\delta,\mathscr{P}}_{\pi}(x) = \underset{\tilde{\mathscr{P}}\in \mathscr{D}^{\delta}}{\text{sup}} \ S_{\pi}^{\tilde{\mathscr{P}}}(x)= \underset{\tilde{\mathscr{P}}\in \mathscr{D}^{\delta}}{\text{sup}} \ \mathds{E}_{\pi}^{{\tilde{\mathscr{P}}}} \Big[ \sum_{t=0}^{\tau-1} \kappa^{{\tilde{\mathscr{P}}}}(X_t, A_t)\big| X_0 = x  \Big]
    \end{split}
  \end{equation*}
  \normalsize
  \end{proof}
 We could also express the robust safety function $S^{\delta,\mathscr{P}}_{\pi}(x)$ in an alternative form.
 \begin{lem} \label{dr_safety_fn_rew}
  The robust safety function $S^{\delta,\mathscr{P}}_{\pi}(x)$ can be expressed as:
  \begin{equation*}
     S^{\delta,\mathscr{P}}_{\pi}(x) = \underset{\tilde{\mathscr{P}}\in \mathscr{D}^{\delta}}{\text{sup}} \ \mathds{E}_{\pi}^{\tilde{\mathscr{P}}} \Big[ \sum_{t=0}^{\tau-1} c_{t+1} \big| X_0=x \Big],
  \end{equation*}
  where,
  \begin{equation*}
      \begin{split}
          c_{t+1}=\begin{cases}
              1, \text{ if } X_{t+1}\in U \\
              0, \text{ otherwise. }
          \end{cases}
      \end{split}
  \end{equation*}
 \end{lem}
 \begin{proof}
     For each $t\in \{0,1,...\}$, we have 
     \begin{equation*}
    \begin{split}
         & \mathds{E}_{\pi}^{\tilde{\mathscr{P}}} [c_{t+1}\big| X_0=x] \\
            & = \mathds{E}_{\pi}^{\tilde{\mathscr{P}}} \big[\mathds{E}^{\tilde{\mathscr{P}}}_{\pi} [c_{t+1}|X_t,A_t]\big| X_0=x\big] \\
            & = \mathds{E}_{\pi}^{\tilde{\mathscr{P}}} [ \kappa^{\tilde{\mathscr{P}}}(X_t,A_t)\big| X_0=x]
    \end{split}
     \end{equation*}
     Hence, 
     \begin{equation*}
    \begin{split}
         & \mathds{E}_{\pi}^{\tilde{\mathscr{P}}} \big[\sum_{t=0}^{\tau-1}  c_{t+1} \big] =  \mathds{E}_{\pi}^{x,\tilde{\mathscr{P}}} \big[\sum_{t=0}^{\tau-1}  \kappa^{\tilde{\mathscr{P}}}(X_t,A_t) \big| X_0=x \big] \overset{(a)}{=}  S^{\delta,\mathscr{P}}_{\pi}(x)
    \end{split}
     \end{equation*}
     Equality $(a)$ follows from Lemma \ref{dr_safe_kappa}.
 \end{proof}

 We notice that the robust safety function $S^{\delta,\mathscr{P}}_{\pi}(x)$ can be thought of as a value function. To this end, we define a distributionally robust Q-function for each state-action pair and relate it with the robust safety function.
 \begin{defn}\label{Q_fn_def} [Robust Q-function]
     Suppose $X_0=x$ and $A_0=a$.
     Then, the distributionally robust Q-function $Q^{\delta,\mathscr{P}}_{\pi}(x,a)$ for each $(x,a)\in H\times \mathcal{A}$ is defined as 
     \begin{equation*}
     \begin{split}
       & Q^{\delta,\mathscr{P}}_{\pi}(x,a) := \underset{\tilde{\mathscr{P}}\in \mathscr{D}^{\delta}}{\text{sup}} \ \mathds{E}_{\pi}^{x,\tilde{\mathscr{P}}} \Big[\sum_{t=0}^{\tau-1} c_{t+1}  \big| X_0=x, A_0=a \Big], 
     \end{split} 
     \end{equation*}
     where, $c_{t+1}$ is as defined in Lemma \ref{dr_safety_fn_rew}. \remend
 \end{defn}
 \par The following result provides an upper bound for the robust safety function.
 \begin{lem}
     The robust safety function can be expressed in terms of the robust Q-function as follows:
     \begin{equation*}
         S^{\delta,\mathscr{P}}_{\pi}(x) \leq \sum_{a\in \mathcal{A}} \pi(a|x) Q^{\delta,\mathscr{P}}_{\pi}(x,a) 
     \end{equation*}
 \end{lem}
 \begin{proof}
     From Lemma \ref{dr_safety_fn_rew} and Definition \ref{Q_fn_def},
     \begin{equation*}
     \begin{split}
     &S^{\delta,\mathscr{P}}_{\pi}(x)\\
     &= \underset{\tilde{\mathscr{P}}\in \mathscr{D}^{\delta}}{\text{sup}} \ \mathds{E}_{\pi}^{\tilde{\mathscr{P}}} \Big[ \sum_{t=0}^{\tau-1} c_{t+1} \big| X_0=x \Big]\\
     & = \underset{\tilde{\mathscr{P}}\in \mathscr{D}^{\delta}}{\text{sup}} \ \sum_{a\in \mathcal{A}} \pi(a|x) \ \big[ \mathds{E}_{\pi}^{\tilde{\mathscr{P}}} \Big[ \sum_{t=0}^{\tau-1} c_{t+1} \big| X_0=x, A_0=a \Big] \big] \\
     & \leq  \sum_{a\in \mathcal{A}} \pi(a|x) \ \underset{\tilde{\mathscr{P}}\in \mathscr{D}^{\delta}}{\text{sup}} \  \mathds{E}_{\pi}^{\tilde{\mathscr{P}}} \Big[ \sum_{t=0}^{\tau-1} c_{t+1} \big| X_0=x, A_0=a \Big]  \\
     & = \sum_{a\in \mathcal{A}} \pi(a|x) Q^{\delta,\mathscr{P}}_{\pi}(x,a)
     \end{split}
  \end{equation*}
 \end{proof}
 \begin{rem}
     We express the robust safety function in terms of the robust Q-function for the following reasons. With the ambiguity set around the nominal $P_{x,a}(y)$, it is much easier and natural to deal with $Q^{\delta,\mathscr{P}}_{\pi}(x,a)$ than with $S^{\delta,\mathscr{P}}_{\pi}(x)$. We will observe that in the convex program formulation of the Q-function.
 \end{rem}
\par We now show that the robust Q-function $Q^{\delta,\mathscr{P}}_{\pi}(.)$ can also be computed recursively using dynamic programming.
 \begin{lem}\label{dr_q_dp}
     The robust Q-function $Q^{\delta,\mathscr{P}}_{\pi}(x,a)$ is given by
     \small
  \begin{equation*}
  \begin{split}
     Q^{\delta,\mathscr{P}}_{\pi}(x,a)  =& \underset{\tilde{\mathcal{P}}_{x,a}\in \mathcal{D}^{\delta}_{x,a}}{\text{sup}}  \mathds{E}_{\pi}^{\tilde{\mathcal{P}}_{x,a}} \Big[c(x,a,y) \\
    &+  Q^{\delta,\mathscr{P}}_{\pi}(y,a') \big| X_0=x, A_0=a, y\sim \tilde{\mathcal{P}}_{x,a}, a'\sim \pi(a'|y) \Big],
     \end{split}
  \end{equation*}
  \normalsize
 where $c(x,a,y)=1$ if $y\in U$, else $c(x,a,y)=0$.
 \end{lem}
\begin{proof}
    Using Lemma \ref{dr_safe_kappa}, we can express the robust Q-function as follows:
    \small
    \begin{equation*}
    \begin{split}
    & Q^{\delta,\mathscr{P}}_{\pi}(x,a) = \underset{\tilde{\mathscr{P}}\in \mathscr{D}^{\delta}}{\text{sup}} \ \mathds{E}_{\pi}^{\tilde{\mathscr{P}}} \sum_{t=0}^{\tau-1} \Big[ \kappa^{\tilde{{\mathscr{P}}}}(X_t, A_t) \big| X_0=x,A_0=a \Big].
    \end{split}  
    \end{equation*}
    \normalsize
    Then, similar to Lemma 2 in \cite{mazumdar2024safe}, we can further express $Q^{\delta,\mathscr{P}}_{\pi}(x,a)$ recursively as:
    \small
    \begin{equation*}
     \begin{split}
     &Q^{\delta,\mathscr{P}}_{\pi}(x,a)\\  &= \underset{\tilde{\mathcal{P}}_{x,a}\in \mathcal{D}^{\delta}_{x,a}}{\text{sup}}  \mathds{E}_{\pi}^{\tilde{\mathcal{P}}_{x,a}} \Big[\kappa^{\tilde{{\mathscr{P}}}}(x,a) \\
    &+  Q^{\delta,{\mathscr{P}}}_{\pi}(y,a') \big| X_0=x, A_0=a, y\sim \tilde{\mathcal{P}}_{x,a}, a'\sim \pi(a'|y) \Big]
     \end{split}   
    \end{equation*}
    \normalsize
    Since, $\mathds{E}_{\pi}^{\tilde{\mathcal{P}}_{x,a}}\big[c(x,a,y)|X_0=x, A_0=a, y\sim \tilde{\mathcal{P}}_{x,a}\big]=\kappa^{\tilde{{\mathscr{P}}}}(x,a)$, we get the desired result.
\end{proof}
 \par Following result demonstrate that the robust Q-function is time-invariant in nature.
 \begin{lem}
   Suppose at a time step $t$, $X_t=x$ and $A_t=a$.
     Then, the robust Q-function $Q^{\delta,\mathscr{P}}_{\pi}(x,a)$ is given by
     \small
  \begin{equation*}
  \begin{split}
     Q^{\delta,\mathscr{P}}_{\pi}(x,a)  = & \underset{\tilde{\mathcal{P}}_{x,a}\in \mathcal{D}^{\delta}_{x,a}}{\text{sup}} \ \mathds{E}_{\pi}^{\tilde{\mathcal{P}}_{x,a}} \Big[c(x,a,y)\\
     &+  Q^{\delta,\mathscr{P}}_{\pi}(y,a) \big| X_t=x, A_t=a, y\sim \tilde{\mathcal{P}}_{x,a}, a'\sim \pi(a'|y) \Big],
     \end{split}
  \end{equation*}
  \normalsize
 where $c(x,a,y)$ as defined in Lemma \ref{dr_q_dp}. 
 \end{lem}
 \begin{proof}
     The proof follows from the fact that the transition probabilities are time invariant.
 \end{proof}

\subsection{Convex program formulation of the robust Q-function and Model-based robust Q-iteration:}
In the following lemma, we express the robust Q-function as a finite-dimensional optimization problem. To this end, we make use of the Kantorovich duality result for Wasserstein distance computation.
\begin{lem} \label{Q_dual}
    For each state-action pair $(x,a)\in \mathcal{X}\times \mathcal{A}$, the distributionally robust Q-function is the solution of:
    \small
    \begin{equation}
    \begin{split}
       & Q^{\delta,\mathscr{P}}_{\pi}(x,a)= \underset{\lambda \geq 0}{\text{inf}} \ \Big[\lambda \delta  \\
        & + \sum_{y\in \mathcal{X}} \ \underset{l\in \mathcal{X}}{\text{max}} \ \Big( -\lambda |l-y| + c(x,a,l) + \sum_{a'\in \mathcal{A}} Q^{\delta,\mathscr{P}}_{\pi}(l,a') \pi(a'|l) \Big) {P}_{x,a}(y) \Big]
    \label{robsut_Q}
    \end{split}
\end{equation}
\normalsize
\end{lem}
\begin{proof}
    From the definition, $Q^{\delta,\mathscr{P}}_{\pi}(x,a)$ is the solution of the above optimization problem
    \small
    \begin{equation}
        \begin{split}
           & Q^{\delta,\mathscr{P}}_{\pi}(x,a) \\
           =  & \underset{{\tilde{\mathcal{P}}_{x,a}}\in \mathcal{D}^{\delta}_{x,a}}{\text{sup}} \mathds{E}^{\tilde{\mathcal{P}}_{x,a}}_{\pi} \big[ c(x,a,y) + Q^{\delta,\mathscr{P}}_{\pi}(y,a') \big| Z^{x,a}_t \big] \\
             & = \begin{cases}
                    \underset{{\tilde{\mathcal{P}}_{x,a}}\in \mathscr{M}(\mathcal{X})}{\text{sup}} \mathds{E}_{\pi}^{\tilde{\mathcal{P}}_{x,a}} \big[ c(x,a,y) + Q^{\delta,\mathscr{P}}_{\pi}(y,a') \big| Z_{t}^{x,a}  \big] \\
                    \text{s.t. } W(\tilde{\mathcal{P}}_{x,a},\mathcal{P}_{x,a}) \leq \delta
                \end{cases}  \\
                & \overset{(a)}{=} \begin{cases}
                    \underset{{\tilde{{P}}_{x,a}}\in \mathscr{M}(\mathcal{X})}{\text{sup}} \mathds{E}_{\pi}^{\tilde{{P}}_{x,a}} \big[ c(x,a,y) +  Q^{\delta,\mathscr{P}}_{\pi}(y,a') \big| Z_t^{x,a}  \big] \\
                    \text{s.t. }   \sum_{z_1\in \mathcal{X}} f(z_1) \tilde{{P}}_{x,a}(z_1) \\
                    + \sum_{z_2\in \mathcal{X}} \underset{z_1}{\text{inf}} \ \big( ||z_1-z_2|| - f(z_1) \big) {P}_{x,a}(z_2) \leq \delta; \ \forall f \in l^1
                \end{cases},  
        \end{split}
    \end{equation}
    \normalsize
    where $Z^{x,a}_t:=\{X_t=x, A_t=a, y\sim \tilde{\mathcal{P}}_{x,a}, a'\sim \pi(a'|y)$\}.\\
In the above expression, equality $(a)$ is due to \eqref{Wasserstein_dual2}. Using the standard duality in constrained optimization, we further get:
\small
    \begin{equation}
        \begin{split}
           & Q^{\delta,\mathscr{P}}_{\pi}(x,a) \\
         & = \underset{\lambda \geq 0}{\text{inf}} \   \underset{{\tilde{\mathcal{P}}_{x,a}}\in \mathscr{M}(\mathcal{X})}{\text{sup }}  \sum_{y\in \mathcal{X}} \big[ c(x,a,y) + \sum_{a'\in \mathcal{A}} Q^{\delta,\mathscr{P}}_{\pi}(y,a') \pi(a'|y) \big] \tilde{{P}}_{x,a}(y) \\
                 &   - \lambda \Big(  \sum_{z_1\in \mathcal{X}} f(z_1) \tilde{{P}}_{x,a}(z_1) + \sum_{z_2\in \mathcal{X}} \underset{z_1}{\text{min}} \ \big( |z_1-z_2| - f(z_1) \big) {P}_{x,a}(z_2) - \delta \Big);\\
                 & \ \forall f \in L^1 \\
        & = \underset{\lambda \geq 0}{\text{inf}}   \underset{{\tilde{\mathcal{P}}_{x,a}}\in \mathscr{M}(\mathcal{X})}{\text{sup }}  \sum_{y\in \mathcal{X}} \big[ c(x,a,y) \\
        & + \sum_{a'\in \mathcal{A}} Q^{\delta,\mathscr{P}}_{\pi}(y,a') \pi(a'|y) - \lambda f(y)  \big] \tilde{{P}}_{x,a}(y) \\
                 &   - \lambda \Big(  \sum_{z_2\in \mathcal{X}} \underset{z_1}{\text{min}} \ \big( |z_1-z_2| - f(z_1) \big) {P}_{x,a}(z_2) - \delta \Big); \ \forall f \in L^1 \\
        & = \underset{\lambda \geq 0}{\text{inf}}   \underset{{\tilde{\mathcal{P}}_{x,a}}\in \mathscr{M}(\mathcal{X})}{\text{sup }}  \sum_{y\in \mathcal{X}} \big[ c(x,a,y) \\
        & + \sum_{a'\in \mathcal{A}} Q^{\delta,\mathscr{P}}_{\pi}(y,a') \pi(a'|y) - \lambda f(y)  \big] \tilde{{P}}_{x,a}(y) \\
                 &   + \lambda \Big(  \sum_{z_2\in \mathcal{X}} \underset{z_1}{\text{max}} \ \big( -|z_1-z_2| + f(z_1) \big) {P}_{x,a}(z_2) + \delta \Big); \ \forall f \in L^1 \\
      &  \overset{a}{=} \underset{\lambda \geq 0}{\text{inf}} \ \Big[\lambda \delta \\
      & + \sum_{y\in \mathcal{X}} \ \underset{l\in \mathcal{X}}{\text{max}} \ \Big( -\lambda |l-y| + c(x,a,l) + \sum_{a'\in \mathcal{A}} \ Q^{\delta,\mathscr{P}}(l,a')\pi(a'|l) \Big) {P}_{x,a}(y) \Big]
        \end{split}
    \end{equation}
    \normalsize
To get equality $(a)$ in the above expression, we use the following analysis: since the expression is valid for all $f\in L^1$, we consider $f(y)$ such that $\lambda f(y)=-(c(x,a,y) + \sum_{a'\in \mathcal{A}} Q^{\delta,\mathscr{P}}_{\pi}(y,a')\pi(a'|y))$ for all $y\in \mathcal{X}$. This choice of $f(y)$ helps us to avoid $\tilde{P}_{x,a}(y)$ thus eliminating the inner optimization. 
\end{proof}
\par As a corollary to the above result, we can further rewrite the robust Q-function as a finite convex program, as presented below.
\begin{cor}
    The robust Q-function $Q^{\delta,\mathscr{P}}_{\pi}(x,a)$, $(x,a)\in H\times \mathcal{A}$, is the optimal value of the following convex program:
    \small
    \begin{equation}
    \begin{split}    
       & \underset{\lambda \geq 0, h(y)\in \mathds{R}}{\text{inf}} \ \Big( \lambda \delta + \sum_{y\in \mathcal{X}} h(y) P_{x,a}(y) \Big) \\
      & \text{s.t.} \  \underset{l\in \mathcal{X}}{\text{max}}  \Big(- \lambda |l-y| + c(x,a,l) + \sum_{a'\in \mathcal{A}} \ Q^{\delta}(l,a') \pi(a'|l) \Big) \leq h(y); \forall y\in \mathcal{X} 
    \end{split}
    \end{equation}
    \normalsize
\end{cor}
\subsection{Distributionally robust Q-iteration}
In order to compute the robust safety function $S^{\delta,\mathscr{P}}_{\pi}(x)$, we need to solve $|H|\times |\mathcal{A}|$ dimensional system of equations of the form \eqref{robsut_Q}. If the dimension of the state and action spaces is large, it will be very inefficient to solve it directly. Instead, similar to standard value iteration, we can compute  $Q^{\delta,\mathscr{P}}_{\pi}(x,a)$ for all $(x,a)\in H\times \mathcal{A}$ by simultaneously updating $Q^{\delta,\mathscr{P}}_{\pi}(x,a)$ from some arbitrary initial value recursively until convergence from some initial $Q^{\delta,\mathscr{P}}_{\pi}(x,a)$. This is the so-called value iteration algorithm. It has been demonstrated in \cite{iyengar2005robust} that in the robust value iteration algorithm, the robust value function (robust Q-function in our case) converges to the true robust value function (true robust Q-function) similar to the standard value iteration algorithm. 
\par We now present a similar algorithm to evaluate the robust Q-function $Q_{\pi}^{\delta,\mathscr{P}}(x,a)$ with the known nominal probability distribution $\mathcal{P}_{x,a}$. In the following pseudo-code, for the ease of presentation, we use $Q^{\delta,\mathscr{P}}(x,a)$ instead of $Q_{\pi}^{\delta,\mathscr{P}}(x,a)$.

\begin{algorithm}
  \caption{: Distributionally robust Q-iteration for estimating the upper bound of robust safety function \( S^{\delta,\mathscr{P}}_{\pi}(x) \)}
  \begin{algorithmic}[1]
    \State \textbf{Input:} Set of states $\mathcal{X}$, set of actions $\mathcal{A}$, nominal transition probabilities $\mathscr{P}$, \( c(x, a, y) \), robustness parameter \( \delta \), evaluation policy $\pi$,  small threshold \( \theta \) for convergence
    \State \textbf{Initialize:} \( Q^{\delta,\mathscr{P}}(x,a) \leftarrow 0 \) for all \( (x,a) \in \mathcal{X} \times \mathcal{A} \)
    \Repeat
        \State \( \Delta \leftarrow 0 \)
        \For{each state \( x \in \mathcal{X} \)}
            \For{each action \( a \in \mathcal{A} \)}
                \State Store current \( Q \) value: \( Q^{\delta,\mathscr{P}}_{\text{old}}(x,a) \leftarrow Q^{\delta,\mathscr{P}}(x,a) \)
                \State Update \( Q \)-value by solving \eqref{robsut_Q}
                \State \( \Delta \leftarrow \max(\Delta, |Q^{\delta,\mathscr{P}}_{\text{old}}(x,a) - Q^{\delta,\mathscr{P}}(x,a)|) \)
            \EndFor
        \EndFor
    \Until{ \( \Delta < \theta \) }

    \For{each state \( x \in \mathcal{X} \)}
        \State Compute \(\sum_{a \in \mathcal{A}} \pi(a|x) Q^{\delta,\mathscr{P}}(x,a) \)
    \EndFor

    \State \textbf{Output:} Upper bound of robust safety function \( S^{\delta,\mathscr{P}}_{\pi}(x) \)
  \end{algorithmic}
\end{algorithm}

\section{Numerical results}\label{numerical results}

We consider an MDP as shown in Fig. 1. with $11$ states $\mathcal{X}=\{1,2,...,11\}$ and $2$ actions $\mathcal{A}=\{1,2\}$, $2$ goal states $E=\{8,10\}$ and $2$ forbidden or unsafe states $U=\{9,11\}$. Each arrow in Fig. 1. corresponds to an action. The nominal transition probabilities  \( P_{x,a}(y) \) are defined as follows:

\[
\begin{aligned}
    &P_{1,1}(2) = 0.4, \quad P_{1,1}(3) = 0.6, \\
    &P_{1,2}(2) = 0.6, \quad P_{1,2}(3) = 0.4, \\
    &P_{2,1}(4) = 0.5, \quad P_{2,1}(5) = 0.5, \\
    &P_{2,2}(4) = 0.7, \quad P_{2,2}(5) = 0.3, \\
    &P_{3,1}(6) = 0.4, \quad P_{3,1}(7) = 0.6, \\
    &P_{3,2}(6) = 0.6, \quad P_{3,2}(7) = 0.4, \\
    &P_{4,1}(8) = 0.5, \quad P_{4,1}(9) = 0.5, \\
    &P_{4,2}(8) = 0.8, \quad P_{4,2}(9) = 0.2, \\
    &P_{5,1}(4) = 0.4, \quad P_{5,1}(8) = 0.6, \\
    &P_{5,2}(4) = 0.6, \quad P_{5,2}(8) = 0.4, \\
    &P_{6,1}(7) = 0.5, \quad P_{6,1}(10) = 0.5, \\
    &P_{6,2}(7) = 0.55, \quad P_{6,2}(10) = 0.45, \\
    &P_{7,1}(10) = 0.7, \quad P_{7,1}(11) = 0.3, \\
    &P_{7,2}(10) = 0.3, \quad P_{7,2}(11) = 0.7.
\end{aligned}
\]

We consider that the evaluation policy $\pi$ is a uniform policy for all states, i.e., $\pi(a|x)=0.5$, and the safety parameter is $p=0.5$. To evaluate robust $p$-safety, we compute the upper bound of the robust safety function $S_{\pi}^{\delta,\mathscr{P}}(x)$ denoted by $\mathscr{J}(x)=\sum_{a\in \mathcal{A}} \pi(a|x) Q^{\delta,\mathscr{P}}_{\pi}(x,a)$. TABLE I demonstrates the upper bound of the robust safety function concerning different $\delta$. From TABLE I, we observe that with an ambiguity radius up to $\delta=0.25$, the MDP is robust $p$-safe. However, for $\delta=0.3$, safety can not be certified.

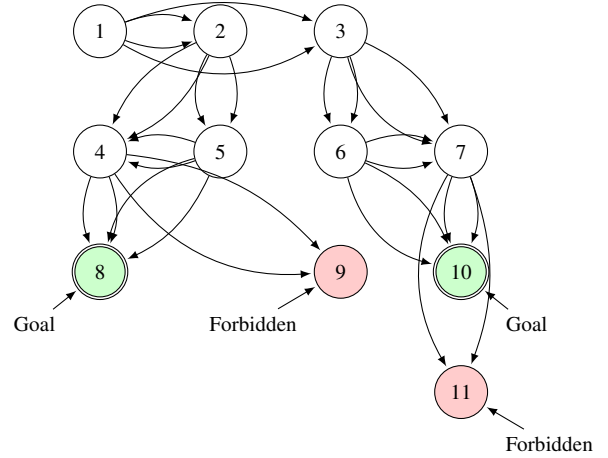
\begin{figure}[h]
    \centering
    \begin{tikzpicture}[->, >=latex, shorten >=1pt, auto, node distance=2cm, scale=0.8, transform shape]

        \node[state] (1) {1};
        \node[state, right of=1] (2) {2};
        \node[state, right of=2] (3) {3};
        \node[state, below of=1] (4) {4};
        \node[state, right of=4] (5) {5};
        \node[state, right of=5] (6) {6};
        \node[state, right of=6] (7) {7};
        
        \node[state, accepting, below of=4, fill=green!20] (8) {8}; 
        \node[state, accepting, below of=7, fill=green!20] (10) {10}; 
        
        \node[state, below of=6, fill=red!20] (9) {9}; 
        \node[state, below of=10, fill=red!20] (11) {11}; 

        \path[->] (1) edge[bend left=20] (2)
                  (1) edge[bend left=20] (3)
                  (1) edge[bend right=20] (2)
                  (1) edge[bend right=30] (3);

        \path[->] (2) edge[bend left=20] (4)
                  (2) edge[bend left=20] (5)
                  (2) edge[bend right=20] (4)
                  (2) edge[bend right=30] (5);

        \path[->] (3) edge[bend left=20] (6)
                  (3) edge[bend left=20] (7)
                  (3) edge[bend right=20] (6)
                  (3) edge[bend right=30] (7);

        \path[->] (4) edge[bend left=20] (8)
                  (4) edge[bend left=20] (9)
                  (4) edge[bend right=20] (8)
                  (4) edge[bend right=30] (9);

        \path[->] (5) edge[bend left=20] (4)
                  (5) edge[bend left=20] (8)
                  (5) edge[bend right=20] (4)
                  (5) edge[bend right=30] (8);

        \path[->] (6) edge[bend left=20] (7)
                  (6) edge[bend left=20] (10)
                  (6) edge[bend right=20] (7)
                  (6) edge[bend right=30] (10);

        \path[->] (7) edge[bend left=20] (10)
                  (7) edge[bend left=20] (11)
                  (7) edge[bend right=20] (10)
                  (7) edge[bend right=30] (11);

        \node[below left=0.3cm and 0.3cm of 8] (goal8) {Goal};
        \draw[->] (goal8) -- (8);

        \node[below right=0.3cm and 0.3cm of 10] (goal10) {Goal};
        \draw[->] (goal10) -- (10);

        \node[below left=0.3cm and 0.3cm of 9] (forbidden9) {Forbidden};
        \draw[->] (forbidden9) -- (9);

        \node[below right=0.3cm and 0.3cm of 11] (forbidden11) {Forbidden};
        \draw[->] (forbidden11) -- (11);

    \end{tikzpicture}
    \caption{Example MDP Diagram with Goal and Forbidden States}
    \label{fig:mdp_diagram}
\end{figure}

\begin{table}[h!]
\centering
\textbf{Values of \(\sum_{a\in \mathcal{A}} \pi(a|x) Q^{\delta,\mathscr{P}}_{\pi}(x,a)\) for different \(\delta\)} 
\vspace{0.3em} 

\scriptsize{%
\begin{tabular}{|c|c|c|c|c|c|c|c|}
\hline
\(\delta\) & \(\mathscr{J}(1)\) & \(\mathscr{J}(2)\) & \(\mathscr{J}(3)\) & \(\mathscr{J}(4)\) & \(\mathscr{J}(5)\) & \(\mathscr{J}(6)\) & \(\mathscr{J}(7)\) \\
\hline
0 & 0.1734 & 0.0800 & 0.2669 & 0.1000 & 0.0500 & 0.1837 & 0.3500 \\
\hline
0.05 & 0.2290 & 0.1301 & 0.3100 & 0.1345 & 0.1017 & 0.2267 & 0.3782 \\
\hline
0.1 & 0.2844 & 0.1826 & 0.3522 & 0.1703 & 0.1555 & 0.2703 & 0.4068 \\
\hline
0.15 & 0.3388 & 0.2371 & 0.3935 & 0.2077 & 0.2115 & 0.3147 & 0.4359 \\
\hline
0.2 & 0.3917 & 0.2933 & 0.4338 & 0.2466 & 0.2698 & 0.3599 & 0.4655 \\
\hline
0.25 & 0.4426 & 0.3509 & 0.4732 & 0.2870 & 0.3304 & 0.4059 & 0.4957 \\
\hline
0.3 & 0.4912 & 0.4095 & 0.5116 & 0.3289 & 0.3934 & 0.4526 & 0.5263 \\
\hline
\end{tabular}%
}

\vspace{0.3em} 
\captionsetup{justification=centering} 
\caption{}
\label{tab:sample_table}
\end{table}

\section{Conclusion and Future Work} \label{sec_conclusion}
We addressed the problem of distributionally robust safety verification for Markov Decision Processes (MDPs). Specifically, we introduced the concept of robust probabilistic safety, termed robust $p$-safety, which generalizes probabilistic safety to MDPs with uncertain transition probabilities. We defined a robust safety function and derived an upper bound for it in terms of a robust Q-function to evaluate robust safety. We then presented a method to compute this robust Q-function by solving a finite convex optimization problem. Building on this, we developed a recursive algorithm based on robust Q-iteration to iteratively compute an upper bound on the robust safety function.
\par Future work will extend this framework to MDPs with larger state-action spaces and potentially continuous state spaces. Additionally, an interesting direction for further research is the development of online algorithms that can perform robust safety verification without reliance on a nominal model.
\addtolength{\textheight}{-12cm}   





%
%
%
%
%
%
\bibliographystyle{IEEEtran}
\bibliography{ecc24}
\end{document}